\documentclass[journal]{IEEEtran}

\usepackage{upgreek}
\usepackage{amsmath} 
\usepackage{float} 
\usepackage{url} 
\usepackage{comment}
\usepackage{graphicx}
\usepackage{tablefootnote}
\usepackage{tabularx,booktabs}
\usepackage{enumitem,booktabs,cfr-lm}
\usepackage[referable]{threeparttablex}
\newcolumntype{C}{>{\centering\arraybackslash}X} % centered version of "X" type
\usepackage[colorlinks,allcolors=blue]{hyperref}
\ifCLASSINFOpdf
\else
\fi

\usepackage{lineno}
\begin{document}

%\linenumbers

\title{Characterization of SiPM Avalanche Triggering Probabilities}

 \widowpenalty10000
  \clubpenalty10000

\author{G.~Gallina, F.~Reti\`ere, P.~Giampa, J.~Kroeger, P.~Margetak, \\ S. Byrne Mamahit, A.~De St. Croix, F.~Edaltafar,  N.~Massacret, M.~Ward, G.~Zhang
\thanks{G. Gallina (giacomo@triumf.ca), F.~Reti\`ere, P.~Giampa, P.~Margetak, F.~Edaltafar, N.~Massacret and A.~De St. Croix  are with TRIUMF, Vancouver, British Columbia V6T 2A3, Canada}
\thanks{G. Gallina and A.~De St. Croix  are also with Department of Physics and Astronomy, University of British Columbia, 6224 Agricultural Road Vancouver BC, CA}
\thanks{J. Kroeger is with Physikalisches Institut der Universit\"at Heidelberg, Heidelberg, Germany}
\thanks{S. Byrne Mamahit is with University of Waterloo, 200 University Ave. W, Waterloo, ON, N2L 3G1}
\thanks{M. Ward is with Queen's University. Department of Physics, 64 Bader Lane, Kingston, ON, Canada, K7L 3N6}
\thanks{G. Zhang is with School of Science, X\'ian Polytechnic University, X\'ian, 710048, China}% <-this % stops a space
\thanks{Manuscript received April XXX, 2019; revised XXX XXX, 2019}}

\markboth{Journal of TRANSACTIONS ON ELECTRON DEVICES,~Vol.~XX, No.~XX, XX~XX}%
{Shell \MakeLowercase{\textit{et al.}}: Bare Demo of IEEEtran.cls for IEEE Journals}


% make the title areadd
\maketitle

\begin{abstract}
Silicon Photo-Multipliers (SiPMs) are detectors sensitive to single photons that are used to detect scintillation and Cherenkov light in a variety of physics and medical-imaging applications. SiPMs measure single photons by amplifying the photo-generated carriers (electrons or holes) via a Geiger-mode avalanche. The Photon Detection Efficiency (PDE) is the combined 
probability that a photon is absorbed in the active volume of the device with a subsequently triggered avalanche. Absorption and avalanche triggering probabilities are correlated since the latter probability depends on where the photon is absorbed. In this paper, we introduce a physics motivated parameterization of the avalanche triggering probability that describes the PDE of a SiPM as a function of its reverse bias voltage, at different wavelengths. This parameterization is based on the fact that in p-on-n SiPMs the induced avalanches are electron-driven in the ultra-violet and near-ultra-violet ranges, while they become increasingly hole-driven towards the near-infra-red range. The model has been successfully applied to characterize two Hamamatsu MPPCs and one FBK SiPM, and it can be extended to other SiPMs. Furthermore, this model provides key insight on the electric field structure within SiPMs, which can explain the limitation of existing devices and be used to optimize the performance of future SiPMs.
\end{abstract}

% Note that keywords are not normally sed for peerreview papers.

% \fi
\IEEEpeerreviewmaketitle

\section{Introduction}
\label{S:Intro}
Thanks to the rapid evolution of signal processing and light source technologies, the field of light detection has significantly advanced over the last 10 years \cite{Gola2019}. Sensors capable of detecting single photons are of critical importance for a wide range of scientific and commercial applications. Silicon Photo-Multipliers (SiPMs) are an emerging and very promising technology that addresses the challenge of sensing, timing and quantifying low-light signals down to the single-photon level. Additionally, in contrast to the widely used Photo-Multiplier Tubes (PMTs), SiPMs are low-voltage powered, suited for operation at cryogenic temperatures and in strong magnetic and electric fields, with also negligible gain fluctuations \cite{Garutti2011}. SiPMs consist of an array of tightly packaged Single Photon Avalanche Diodes (SPADs) operated above breakdown voltage, in order to generate Geiger-mode avalanches. A key parameter of SiPMs is their Photon Detection Efficiency (PDE), which is defined as the probability for a single photon to produce a detectable current (or charge) pulse. Experimentally, this quantity can be measured as the ratio between the number of photons producing detectable pulses and the total number of photons impinging onto the SiPM surface, usually measured with a reference detector \cite{Gallina2019}. In previous studies the PDE was parameterized as \cite{Zappala2016}
\begin{equation}
\text{PDE}= \text{FF}\cdot\epsilon(\lambda)\cdot\text{T}_{P}(V,\lambda)
\label{eq:PDE_ZAPPALA}
\end{equation}
where FF is the Fill Factor, $i.e.$ the ratio of the sensitive to total area of the device; $\epsilon(\lambda)$ is the device quantum efficiency, $i.e.$ the probability for a photon impinging on the SiPM surface with wavelength $\lambda$ to be transmitted into the silicon, absorbed, and finally converted into an electron-hole pair; $\text{T}_{P}(V,\lambda)$ is the avalanche triggering probability, $i.e.$ the probability for the generated electron-hole pair to initiate a Geiger-mode avalanche inside the depletion layer. Eq. \ref{eq:PDE_ZAPPALA} is, however, an approximation that can only be applied to a limited number of SiPMs \cite{Zappala2016}. In this paper, we propose a new formulation of the SiPM PDE that accounts for the position of photon-absorption. This new parametrization predicts the PDE as a function of the reverse bias voltage and wavelength, corresponding to attenuation lengths in silicon between a few nanometers and several tens of micrometers, specifically accounting for the transition from electron-driven avalanches (close to the surface), to hole-driven avalanches (deeper inside the silicon) in p-on-n SiPMs. The  model  has  been  successfully  applied  to  characterize the response of three SiPMs: two Hamamatsu MPPCs and one Fondazione-Bruno-Kessler (FBK) SiPM, and it can be extended to any other SiPM including n-on-p devices.
%%%%%%%%%%%%%%%%%%%%%

\section{Parametrization of the SiPM PDE}

\subsection{Model for Single Photon Avalanche Diodes}
\label{S:Modellization_intro}

\begin{figure}[h]            
  \centering
  \includegraphics[width=0.99\linewidth]{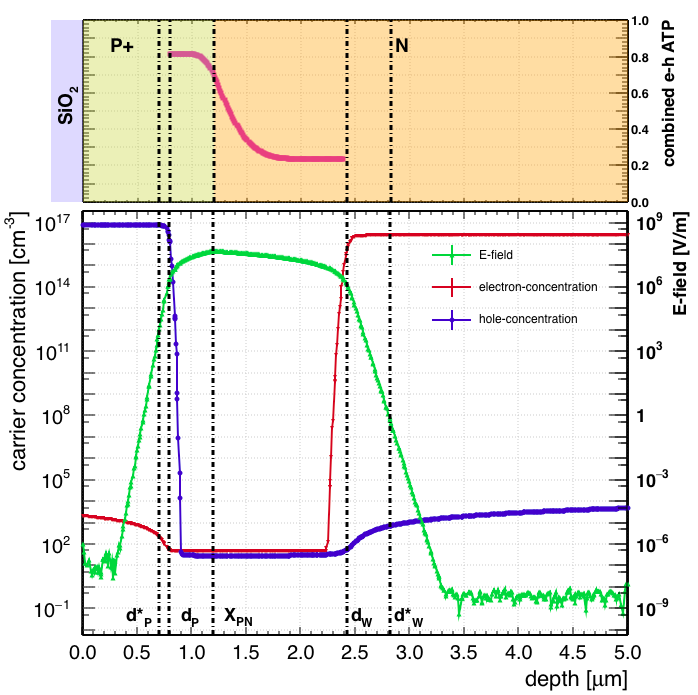}
  \caption{p-on-n SPAD simulated  with Lumerical DEVICE \cite{Lumerical}. The SPAD has an asymmetric constant doping concentration of $\text{N}_\text{P+}= 7.5\cdot 10^{16}\text{ }\text{cm}^{-3}$ and $\text{N}_\text{N}= 2.5\cdot 10^{16}\text{ }\text{cm}^{-3}$. The breakdown voltage for this configuration is $V_{BD}=36.3$ V. Top: Combined electron-hole Avalanche Triggering Probability (ATP): $\text{P}_\text{P}(x,V)$, within the depletion layer for $3.7$ V of over voltage \cite{W.G.1972}. Bottom: Carrier (electron-hole) concentration and electric field profile. $d_P$ and $d_W$ mark the edges of the depletion layer. Additionally, $x_{PN}$ is the position of maximum electric field, while $d_P^*$ and $d_W^*$ are the edges of the effective photon collection region. Different factors like carrier mobility and recombination time \cite{S.M.Sze1995} contribute to defining the exact location of $d_P^*$ and $d_W^*$.}
  \label{F:PNJunction}
\end{figure}

SiPMs are arrays of SPADs separated by guard rings and other structures, such as trenches to suppress optical cross-talk \cite{Piemonte2016}. The field at the edge of each SPAD is expected to be distorted by the proximity of the isolation structures \cite{Acerbi2018}. Nevertheless, the small fraction of the SPAD area affected by 
edge effects and the uniformity of the electric field in the SPAD depletion layer allows to treat the parametrization of the SiPM PDE as a one-dimensional problem \cite{Gulinatti2009}. This approximation may not apply to high-density SiPMs with very small SPADs \cite{Acerbi2018}. Each SiPM SPAD is a reversely biased p-n junction, operated above breakdown. In this configuration, a photo-generated carrier (electron or hole) entering the depletion layer may trigger an avalanche \cite{W.G.1972}.
%and, accordingly to the model, a microplasma biased above the breakdown voltage may be triggered %into avalanche breakdown by a carrier (electron or hole) in the depletion layer \cite{W.G.1972}.
Not every carrier, however, will induce one. Carriers can travel undisturbed or lose energy by interacting with the lattice, recombining before the junction enter in Geiger breakdown \cite{Mazzillo2008}. Additionally, electrons (holes) may be lost if they diffuse into the silicon surface (substrate). One can therefore associate a finite probability of triggering an avalanche to each carrier depending on the SPAD reverse bias voltage $V$ and on the position $x$ in which the carrier enters or is generated in the depletion layer: $\text{P}_e(x,V)$ or $\text{P}_h(x,V)$ \cite{McIntyre1973}. Fig. \ref{F:PNJunction} shows the electrical properties of a typical p-n junction for a p-on-n SiPM simulated with the Lumerical DEVICE simulation package \cite{Lumerical}. The electric field has a maximum at the p-on-n transition labeled $x_{PN}$. The depletion layer starts at $d_P$ (on the P+ side) and ends at $d_W$ (on the N side), defining the total junction length of $W\equiv(d_W-d_P)$ \cite{Otte2018}. Fig. \ref{F:PNJunction} also shows the combined avalanche triggering probability $\text{P}_{\text{P}}(x,V)$ that an electron-hole pair will trigger an avalanche within the depleted region
\begin{equation*}
\text{P}_{\text{P}}(x,V)\equiv \Big(\text{P}_e(x,V)+\text{P}_h(x,V)-\text{P}_e(x,V)\cdot \text{P}_h(x,V)\Big)   
\end{equation*}
This probability is electron-driven at $x=d_P$ and hole-driven at $x=d_W$ yielding $\text{P}_{\text{P}}(d_P,V)=\text{P}_e(d_P,V)$ and  $\text{P}_{\text{P}}(d_W,V)=\text{P}_h(d_W,V)$, respectively \cite{W.G.1972}. The probability is smaller on the N side due to the significantly lower impact-ionization-coefficient for holes compared to electrons \cite{Miller1957}. Carriers created outside of the depleted region may contribute to the total PDE reaching the depletion layer by drifting or diffusing and subsequently triggering an avalanche. In this case, the probability that a carrier reaches the depleted region depends on its lifetime and on the original position of photo-generation \cite{Mazzillo2008}. 

%%The transport probabilities $\text{P}_\text{tr-e}(x)$ and $\text{P}_\text{tr-h}(x)$ can be computed exactly if the p-n junction structure is known (e.g \cite{Gulinatti2009,Xu2016}).
%___________________________________________________________________________

\subsection{Modelling photon absorption and carrier transport}
\label{S:Modellization}

The probability that a single photon of wavelength $\lambda$ is absorbed between $x$ and $x+dx$ as a function of the photon attenuation length $\mu(\lambda)$ is:
\begin{equation}
d\text{P}_{\text{Abs}}(x)=\frac{1}{\mu}\exp{\Bigg(-\frac{x}{\mu}\Bigg)}\text{ }\text{dx}
\label{eq:absorbtion}
\end{equation}
The photon absorption results in the generation of one or more electron-hole pairs \cite{Canfield1998}. Depending on the photon attenuation length and on the location and extension of the microcell depletion layer, the avalanche process can be reduced to one of these three independent mechanisms\cite{Kindt1999}:
\begin{enumerate}
    \item The photon is absorbed in the quasi-neutral upper layer ($x\in[0,d_P]$). The photo-generated electron diffuses (or drifts) to the depleted region triggering an avalanche with probability $\text{P}_\text{tr-e}(x)\cdot\text{P}_e(d_P,V)$, where $\text{P}_\text{tr-e}(x)$ is the probability for the electron produced at $x$ to reach the upper depletion layer boundary $d_P$.
    \item The photon is absorbed in the depleted layer ($x\in[d_P,d_W]$). The photo-generated electron-hole pair triggers an avalanche with probability $\text{P}_{\text{P}}(x,V)$.
    \item  The photon is absorbed in the quasi-neutral lower layer ($x \ge d_W$). The photo-generated hole diffuses (or drifts) to the depleted region triggering an avalanche with probability $\text{P}_\text{tr-h}(x)\cdot\text{P}_h(d_W,V)$, where $\text{P}_\text{tr-h}(x)$ is the probability for the hole produced at $x$ to reach the lower depletion layer boundary $d_W$.
\end{enumerate}

The second process is the dominant mechanism for the avalanche breakdown \cite{Gulinatti2011}. Nevertheless, the drift and diffusion of  minority carriers from the neutral regions can influence the total PDE by producing significant delays in the avalanche generation, as shown experimentally in \cite{Boone2017} and numerically in \cite{Gulinatti2009}.  The data reported in this paper are not directly sensitive to time delays in the avalanche build up (Sec. \ref{S:Model_validation}). Diffusing or drifting electrons (holes) in particular have the same probability of creating avalanches once they enter the depletion layer, regardless of their original creation depth, $i.e.$ $\text{P}_e(d_P,V)$ ($\text{P}_h(d_W,V)$). We can therefore simplify the transport of the photo-generated carriers in the depleted region by introducing two effective depth parameters: $d_P^*$ and $d_W^*$, such that electrons (holes) photo-generated between $d_P^*$ and $d_P$ ($d_W$ and $d_W^*$) always reach the depletion layer boundaries at $d_P$ ($d_W$) (Fig. \ref{F:PNJunction}). With this simplification the electron and hole transport probabilities  $\text{P}_\text{tr-e}(x)$ and $\text{P}_\text{tr-h}(x)$ become step functions as follows:
\begin{equation}
\label{eq:Ptre}
\text{P}_\text{tr-e}(x)=
\begin{cases} 
& 1\text{ if } x\in[d_P^*,d_P]\\
& 0 \text{ else}
\end{cases}
\end{equation}

\begin{equation}
\label{eq:Ptrh}
\text{P}_\text{tr-h}(x)=
\begin{cases} 
& 1\text{ if } x\in[d_W,d_W^*]\\
& 0 \text{ else}
\end{cases}
\end{equation}

%%%%%%%%%%%%%%%%%%%%%

\subsection{Modelling the probability of triggering avalanches within the junction}
\label{S:analytical}

The SiPM PDE for a wavelength $\lambda$ can be obtained by combining the probability of photon absorption (Eq. \ref{eq:absorbtion}) with the simplified transport probabilities (Eq. \ref{eq:Ptre} and Eq. \ref{eq:Ptrh}) and with the combined electron-hole avalanche triggering probability $\text{P}_{\text{P}}(x,V)$ as: 
\begin{equation}
\label{eq:PDE_1}
\text{PDE}=\epsilon_0\cdot\int_{d_P^*}^{d_W^*}\frac{1}{\mu}\exp{\Bigg(-\frac{x}{\mu}\Bigg)}\cdot \text{P}_{\text{P}}(x,V)\text{ }\text{dx}
\end{equation}
where $\epsilon_0$ is the optical efficiency, $i.e.$ the probability that a photon is transmitted in the silicon. This quantity depends on the SPAD fill factor (Sec. \ref{S:Intro}) and reflectivity \cite{Otte2016}. It is worth noting that Eq. \ref{eq:PDE_1} assumes a quantum yield of 1 \cite{Canfield1998}, justified by the range of wavelengths analyzed in Sec. \ref{S:Model_validation}. Eq. \ref{eq:PDE_1} includes all the information necessary to describe the voltage and wavelength dependence of the SiPM PDE. In particular, besides the reduced dimensionality of the problem (Sec. \ref{S:Modellization_intro}), it involves no approximation for the SPAD depletion layer structure ($e.g$ doping profile). The main drawback of Eq. \ref{eq:PDE_1} is that it cannot be expressed using only measurable quantities. The combined electron-hole avalanche triggering probability $\text{P}_{\text{P}}(x,V)$, for example, can be calculated numerically by solving a set of differential equations that depend on the generally unknown SPAD electric field \cite{W.G.1972}. Therefore, an expression of Eq. \ref{eq:PDE_1}, suitable for SiPMs characterization, requires a second approximation in addition to the effective model of the quasi-neutral regions introduced in Sec. \ref{S:Modellization}. Precisely, the avalanche triggering probability $\text{P}_{\text{P}}(x,V)$ is simplified with a step function considering its asymptotic values at the microcell depletion layer boundaries such that\footnote{In the next sections, to keep the notation simple, we will drop the voltage dependence of these quantities.}
\begin{equation}
\label{eq:pp}
\text{P}_{\text{P}}(x)\sim
\begin{cases} 
\text{P}_e(d_P) &\mbox{if } x\in[d_P^*,x_{PN}] \\ 
\text{P}_h(d_W) & \mbox{if } x\in[x_{PN},d_W^*] \\
\end{cases}
\end{equation}
With Eq. \ref{eq:pp}, Eq. \ref{eq:PDE_1} can be integrated exactly obtaining:
\begin{equation}
\label{eq:PDE_final}
\text{PDE}=\text{PDE}_{\text{MAX}}\cdot\Bigg(\text{P}_e(d_P)\cdot f_e^*+\text{P}_h(d_W)\cdot (1-f_e^*)\Bigg)
\end{equation}
where
\begin{equation}
\label{eq:PDEmax}
\text{PDE}_{\text{MAX}}\equiv\epsilon_0\exp{\Bigg(-\frac{d_P^*}{\mu}\Bigg)} \Bigg(1-\exp{\Bigg(-\frac{W^*}{\mu}\Bigg)}\Bigg)
\end{equation}
with $W^*\equiv \Big(d_W^*-d_P^*\Big)$, the length of the effective region in which an absorbed photon can initialize an avalanche process\footnote{$W^*\ge W$ due to the extended junction boundaries (Sec. \ref{S:Modellization}).}. $\text{PDE}_{\text{MAX}}$ represents the saturation PDE for a wavelength $\lambda$ and it is defined as the product of three quantities: (i) the optical efficiency (Sec. \ref{S:analytical}), (ii) the probability that a photon is transmitted through the upper quasi neutral layer, and (iii) the probability that a photon is absorbed in $W^*$.
\begin{equation}
\label{eq:fe}
 f_e^*\equiv \left[\frac{1-\exp{\Big(-\frac{(x_{PN}-d_P^*)}{\mu}\Big)}}{1-\exp{\Big(-\frac{W^*}{\mu}\Big)}} \right]\in[0-1]
\end{equation}
represents the fraction of electron-driven avalanches for a wavelength $\lambda$. It depends on $\big(x_{PN}-d_P^*\big)$: the  length  of  the  region in which avalanches are triggered by an electron.  Considering the weak voltage dependence of $\big(x_{PN}-d_P^*\big)$ and $W^*$ \cite{S.M.Sze1995}, $\text{PDE}_{\text{MAX}}$ and $ f_e^*$  will be considered as voltage-independent quantities. Eq. \ref{eq:PDE_final} offers a compact analytical model for parameterizing the SiPM PDE. However, it depends on two unknown voltage dependent quantities: $\text{P}_e(d_P)$ and $\text{P}_h(d_W)$. 
 
 \subsection{Inferring the electron-hole avalanche triggering probabilities}
 \label{S:PDEUV}
 
The evaluation of $\text{P}_e(d_P)$ and $\text{P}_h(d_W)$ is a complicated numerical problem \cite{McIntyre1973}. Inspection of Eq. \ref{eq:PDE_final} shows an elegant way to find $\text{P}_e(d_P)$ experimentally, without the need to know the microcell electric field. In particular, if the attenuation length for a given wavelength is such that $f_e^*\sim 1$, then Eq. \ref{eq:PDE_final} reduces to
\begin{equation}
\text{PDE}\sim\text{PDE}_{\text{MAX}}\cdot \text{P}_e(d_P)
\label{eq:PDE_UV_exp}
\end{equation}
In this case, the shape of the PDE as a function of the reverse bias voltage simply reduces to the shape of $\text{P}_e(d_P)$, since $\text{PDE}_{\text{MAX}}$ is voltage independent (Sec. \ref{S:analytical}). The condition $f_e^*\sim 1$ for a p-on-n microcell is well verified for UltraViolet (UV) wavelengths due to their attenuation lengths \cite{green_self-consistent_2008}. More generally, the condition $f_e^*\sim 1$ could be also verified for longer wavelengths depending of the junction structure. We propose an equation to overcome the difficulty to interpolate the PDE of UV wavelengths as a function of the reverse bias voltage $V$. Following an approach similar to the one developed in \cite{biroth}, $\text{P}_e(d_P)$ can be expressed as
\begin{equation}
\text{P}_e(d_P) =\left[1-{\Big(k_e\cdot V\cdot \exp{\Big(-k_{e2}/\sqrt{V}\Big)}\Big)}^{-2}\right] 
\label{eq:PDE_UV}
\end{equation}
where $k_e$ and $k_{e2}$ are two voltage-independent parameters used empirically to reproduce the shape of the UV PDE. The problem in evaluating $\text{P}_h(d_W)$ can be solved by introducing a parameter $\text{k}$ that represents an effective ratio of the impact-ionization-coefficients \cite{McIntyre1973}. In this way $\text{P}_h(d_W)$ can be derived from $\text{P}_e(d_P)$ as follows
\begin{equation}
\label{eq:PH}
\text{P}_h(d_W)=\left[1-\Big(1-\text{P}_e(d_P) \Big)^\text{k}\right]
\end{equation}
This equation allows to express Eq. \ref{eq:PDE_final} in term of $\text{P}_e(d_P)$ only, a quantity that can be measured with UV light in p-on-n SiPMs. 
\begin{comment}
We conclude this section noting that the knowledge of  $\text{P}_e(d_P)$ and k can additionally be used to estimate the ionization integral $\delta\equiv\int_0^W\alpha_e(x)\text{ }\text{dx}$ ($i.e.$ the average number of ionizations per carrier pair), solving the following equation \cite{McIntyre1973} :
\begin{equation}
   e^{(1-\text{k})\delta}=\frac{\text{P}_e(d_P)}{(1-\text{P}_e(d_P))}\Bigg[\frac{(1-\text{P}_e(d_P))^\text{k}}{1-(1-\text{P}_e(d_P))^\text{k}}\Bigg]
\end{equation}
The major contribution to the integrand of $\delta$ occurs at fields close to maximum electric field \cite{McIntyre1961}, therefore differently from $\text{k}$ that can  be used only to size the magnitude of the electric field in  the depletion layer ($i.e.$ $\text{k}$ depends only from $E$, Sec. \ref{S:analytical}), $\delta$ additionally accounts also for the extension of the high field region. Therefore, as it will be showed in Sec. \ref{S:Delta_section}, $\text{k}$ and $\delta$ can be used as two independent parameters to compare the effects on the PDE of different electric field configurations.
\end{comment}

\subsection{Data analysis procedure}
\label{S:Summary}

In summary, the procedure to characterize the SiPM PDE for p-on-n junction structures and for different wavelengths is:
\begin{itemize}
    \item Measure $\text{P}_e(d_P)$ by fitting the SiPM PDE as a function of the reverse bias voltage using short wavelengths ($e.g.$ UV) and Eq. \ref{eq:PDE_UV_exp} (combined with Eq. \ref{eq:PDE_UV}). Eq. \ref{eq:PDE_UV_exp} has three voltage-independent fitting parameters: $k_e$, $k_{e2}$, and $\text{PDE}_{\text{MAX}}$. 
    \item  Fit the SiPM PDE as a function of the reverse bias voltage for the other available wavelengths with Eq. \ref{eq:PDE_final} and a unique fit minimization. Eq. \ref{eq:PDE_final} has four fitting parameters $\big(x_{PN}-d_P^*\big)$, $\text{k}$, $W^*$ and $\text{PDE}_{\text{MAX}}$. The first three parameters are wavelength and voltage independent, while $\text{PDE}_{\text{MAX}}$ is a voltage-independent but wavelength-dependent parameter.
\end{itemize}

\section{Model validation}
\label{S:Model_validation}

In this section we will apply the model introduced in Sec. \ref{S:analytical} to characterize the voltage and wavelength dependence of the PDE of three SiPMs with the same junction structure (p-on-n).
%\footnote{The common junction structure is evident comparing Fig. \ref{F:PDEH2017} and Fig. \ref{F:PDEHAMA}. The PDE of UV and near UV wavelengths shows a faster saturation if compared with the PDE of infrared wavelengths. This behaviour is therefore compatible with Eq. \ref{eq:PDE_UV_exp} and opposite to what showed in \cite{W.G.1972} for n-on-p junction structures.}. 
The first SiPM is a Hamamatsu H2017 Multi-Pixel Photon Counter (MPPC) whose characterization was reported in \cite{Girard2018} (Sec. \ref{S:H2017}). The other two SiPMs are: (i) a Hamamatsu VUV4 MPPC (S/N: S13370-6152) and (ii) a FBK Low Field (LF) SiPM (Sec. \ref{S:nexoSiPM}). The characterization of the latter two devices was performed by the nEXO collaboration in \cite{Gallina2019} and \cite{Ako}, mainly for application in liquid xenon. 
 \begin{figure}[h]            
  \centering
  \includegraphics[width=0.99\linewidth]{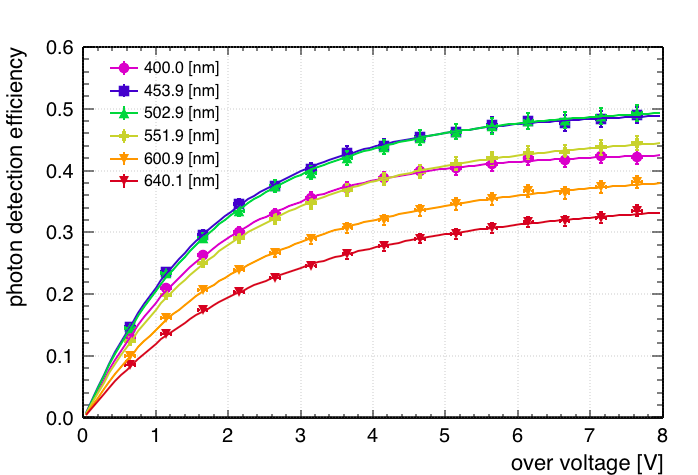}
  \caption{Absolute PDE as a function of the over voltage for the Hamamatsu H2017 MPPC for wavelengths in the range  [400-640] nm. The complete data-set, reported in \cite{Girard2018}, accounts for a wavelength scan with a resolution of $\sim 5$ nm (here displayed every 50 nm).  The solid lines represent fits performed with Eq. \ref{eq:PDE_UV_exp} (for the 400 nm data), and the combined fit with Eq. \ref{eq:PDE_final} (for the other wavelengths).}
  \label{F:PDEH2017}
\end{figure}
The PDE of the Hamamatsu H2017 MPPC was analyzed using the data reported in \cite{Girard2018} that account for wavelengths in the range [400-640] nm. The two nEXO SiPMs were tested in this work using the following wavelengths:  378 nm, 444 nm and 782 nm. The experimental setup, the Data AcQuisition (DAQ) system and the experimental technique used to measure the SiPMs PDE are described in \cite{Gallina2019}. The three wavelengths were obtained with a Hamamatsu Photonics laser controller C10196 run at 500 Hz with PLP-10 laser heads. The Hamamatsu H2017 data were collected at 23.6 $\text{ }\text{C}^\circ$ while the data of the Hamamatsu VUV4 and the FBK LF were recorded at $-40\text{ }\text{C}^\circ$ and $-60\text{ }\text{C}^\circ$, respectively. The dependence of the attenuation length from the temperature was accounted considering the temperature coefficients reported in \cite{green_self-consistent_2008}.

\subsection{Analysis of the Hamamatsu H2017 MPPC}
\label{S:H2017}

\begin{figure}[h]            
  \centering
  \includegraphics[width=0.99\linewidth]{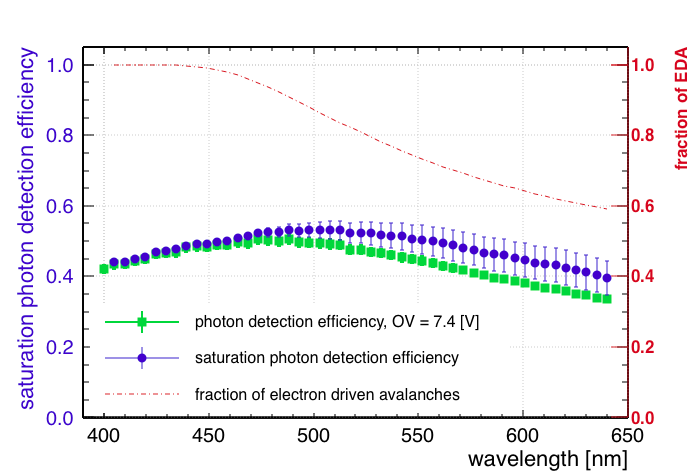}
  \caption{Fraction of Electron Driven Avalanches (EDA) ($f_e^*$, derived using Eq. \ref{eq:fe} and the fit results reported in Table \ref{T:1}) and saturation PDE ($\text{PDE}_{\text{MAX}}$) plotted as a function of the wavelength for the H2017 Hamamatsu MPPC. Due to the slower saturation of the hole probability ($\text{P}_h(d_W)$) compared to the electron one ($\text{P}_e(d_P)$) (see also Fig. \ref{F:PDEPEPHCOMP}), the error on the saturation PDE increases with increasing wavelength. For comparison, in this figure is also reported the photon detection efficiency at the highest Over Voltage (OV) point of Fig. \ref{F:PDEH2017}.}
  \label{F:PDEH2017fe}
\end{figure}

\begin{table*}[!t]
 \caption{Comparison of the parameters derived for the three tested SiPMs}
\label{T:1}
\begin{tabularx}{\textwidth}{@{}l*{10}{C}c@{}}
\toprule
Device          &  $\big(x_{PN}-d_P^*\big)\text{ }[\upmu\text{m}]$ & $W^*\text{ }[\upmu\text{m}]$ & $\text{C}\text{ }[\text{fF}]$ & $W[\upmu\text{m}]$ & $\text{k}$ \\ 
\midrule
Hamamatsu H2017 \cite{Girard2018}   & $1.8\pm 0.1$             & $4.1\pm 0.4$   & $163\pm 1$    & $1.54\pm 0.01$ & $0.25\pm 0.06$  \\  
Hamamatsu VUV4  \cite{Gallina2019}   & $0.8\pm0.2$             & $3.9\pm0.8$   & $116\pm 6$    & $1.01\pm 0.05$ & $0.07\pm 0.06$      \\
FBK LF   \cite{Ako}       & $0.145\pm0.01$            & $2.2\pm0.1$   & $83\pm5$             & $0.92\pm0.06$ & $0.05\pm 0.01$      \\ 
\bottomrule
\end{tabularx}
 \begin{tablenotes}
 \item $\big(x_{PN}-d_P^*\big)$ represents the length of the region in which avalanches are triggered by an electron. $W^*$ is the length of the effective region in which an absorbed photon can initialize an avalanche process. $W$ is the physical junction length derived using: (i) the pixel size and the fill factor provided by each manufacturer, (ii) the single cell capacitance C extrapolated from the SiPM gain \cite{Boone2017}. k is an effective ratio of the impact-ionization-coefficients as reported in \cite{McIntyre1973}.
    \end{tablenotes}
\end{table*}
The absolute PDE of the Hamamatsu H2017 as a function of the over voltage\footnote{The over voltage is defined as the difference between the reverse bias voltage and the breakdown voltage.}, is reported for different wavelengths in Fig. \ref{F:PDEH2017}. The solid lines represent fits performed with Eq. \ref{eq:PDE_UV_exp} and Eq. \ref{eq:PDE_final}. First, the 400 nm data ($\mu(400\text{ nm})\sim 0.1\text{ }\upmu\text{m}$) were fitted with Eq. \ref{eq:PDE_UV_exp} (combined with Eq. \ref{eq:PDE_UV}) to constrain $\text{P}_e(d_P)$, as described in Sec. \ref{S:PDEUV}. The other wavelengths were then fitted all together using Eq. \ref{eq:PDE_final}, with the $\text{P}_e(d_P)$ constrained by the $400\text{ nm}$ fit, and $\big(x_{PN}-d_P^*\big)$, $\text{k}$, $W^*$ and $\text{PDE}_{\text{MAX}}$ as free parameters. The measured effective value for $\text{k}_{\text{H2017}}=0.25\pm 0.06$. The width of the e-triggered avalanche layer is $1.8\pm 0.1\text{ }\upmu\text{m}$ and the effective junction length is $W^*=4.1\pm 0.4\text{ }\upmu\text{m}$. This last quantity can be compared with the physical junction length $W=1.54\pm 0.01 \text{ }\upmu\text{m}$. $W^*$ is bigger than $W$ in agreement with the effect of carrier drift and diffusion described in Sec. \ref{S:analytical}. Additionally, $\text{PDE}_{\text{MAX}}$  and $f_e^*$ (this last quantity extrapolated using Eq. \ref{eq:fe}) are reported in Fig. \ref{F:PDEH2017fe} as a function of the wavelength. $f_e^*$ represents the fraction of electron driven avalanches (Sec. \ref{S:analytical}). It decreases with increasing wavelength which reflects the fact that longer wavelengths are absorbed deeper in the microcell (closer to the N side) and a considerable contribution to the total PDE thus comes from hole-driven avalanches.

\begin{comment}
 \begin{figure}[h]            
  \centering
  \includegraphics[width=0.99\linewidth]{PDE_and_PDEmax_vs_WL_H2017_All.png}
  \caption{Saturation PDE ($i.e.$ $\text{PDE}_{\text{MAX}}$, Eq. \ref{eq:PDE_final}) plotted as a function of the wavelength for the H2017 Hamamatsu MPPC.  For comparison, in this figure is also reported the absolute PDE for the highest Over Voltage (OV) point of Fig. \ref{F:PDEH2017}.}
  \label{F:PDEH2017fe}
\end{figure}
\begin{figure}[h]            
  \centering
  \includegraphics[width=0.99\linewidth]{ATP_vs_OV_H2017.png}
  \caption{Conditional Avalanche Triggering Probability (ATP) as a function of the over voltage for the H2017 Hamamatsu MPPC. The solid lines represent fits performed with Eq. \ref{eq:PDE_UV_exp} (for the 400 nm data) and with Eq. \ref{eq:PDE_final} (for the other wavelengths), accordingly to Sec. \ref{S:Summary}.}
  \label{F:PDEH2017Rescaled}
\end{figure}
\end{comment}

\subsection{Analysis of the SiPMs tested for nEXO}
\label{S:nexoSiPM}

\begin{figure}[h]            
  \centering
  \includegraphics[width=0.99\linewidth]{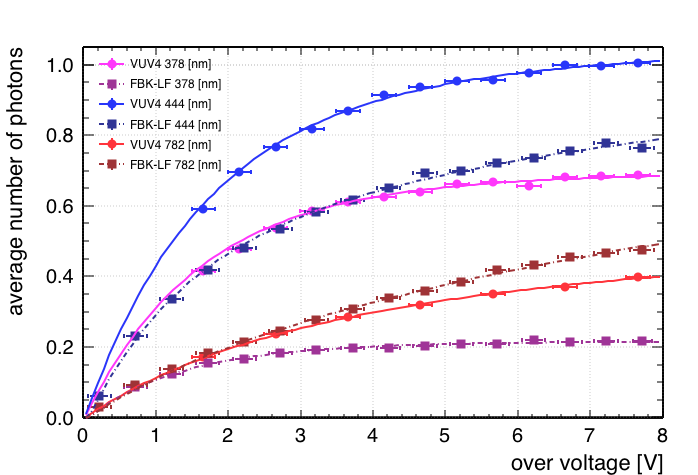}
  \caption{Average number of photons detected by the Hamamatsu VUV4 MPPC and the FBK LF SiPM as a function of the over voltage for not calibrated light fluxes. The  solid and dashed lines represent the fits performed with Eq. \ref{eq:PDE_UV_exp} (for the 378 nm data) and the combined fit with Eq. \ref{eq:PDE_final} (for the other two wavelengths).}
  \label{F:PDEHAMA}
\end{figure}

The same analysis as for the H2017 SiPM was applied to two SiPMs tested for nEXO: the Hamamatsu VUV4 MPPC and the FBK LF SiPM.  In this case we were interested only in relative changes of the shape of the PDE for different wavelength therefore the absolute light fluxes were not calibrated. In Fig. \ref{F:PDEHAMA} we report the average number of photons detected for these two SiPMs. The solid and dashed lines represent fits performed with Eq. \ref{eq:PDE_UV_exp} for the 378 nm data ($\mu(378\text{ nm})\sim 0.03\text{ }\upmu\text{m}$). The other two wavelengths were fitted together using Eq. \ref{eq:PDE_final} and the four free parameters introduced in Sec. \ref{S:Summary}. The effective k values derived from the fit for the FBK LF and the Hamamatsu VUV4 are $\text{k}_{\text{LF}}= 0.05\pm 0.01$ and $\text{k}_{\text{VUV4}}= 0.07\pm 0.06$, respectively. For the FBK LF (Hamamatsu VUV4) the width of the e-triggered avalanche layer is $0.145\pm0.01\text{ }\upmu\text{m}$ ($0.8\pm0.2\text{ }\upmu\text{m}$) and the effective junction length is $2.2\pm0.1\text{ }\upmu\text{m}$ ($3.9\pm0.8\text{ }\upmu\text{m}$). The effective junction of the Hamamatsu VUV4 is less symmetric than the one of the Hamamatsu H2017. Instead the FBK LF has a smaller electron dominated thickness suggesting a stronger doping asymmetry. Additionally, the physical junction length of the  Hamamatsu VUV4 (FBK LF) is $1.01\pm \text{ }0.05\text{ }\upmu\text{m}$ ($0.92\pm0.06\text{ }\upmu\text{m}$). Both these lengths are smaller than the corresponding effective ones and again compatible with the model described in Sec. \ref{S:analytical}. In Table \ref{T:1} we report a summary of the fit parameters for the three SiPMs.
  \begin{figure}[h]            
  \centering
  \includegraphics[width=0.99\linewidth]{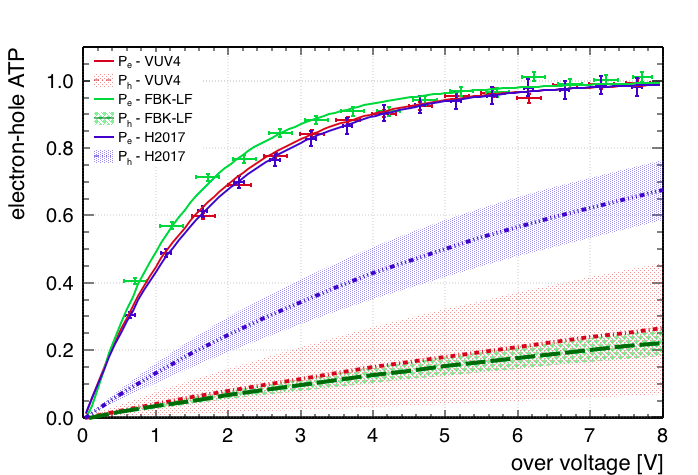}
  \caption{Electron and hole Avalanche Triggering Probability (ATP): $\text{P}_e(d_P)$ and $\text{P}_h(d_W)$, for the three SiPMs analyzed in this paper. For each $\text{P}_e(d_P)$ also are reported the data of Fig. \ref{F:PDEH2017} and Fig. \ref{F:PDEHAMA} used to obtain the corresponding curve. Each $\text{P}_h(d_W)$ is derived from the corresponding $\text{P}_e(d_P)$ using Eq. \ref{eq:PH} and the derived k. The colored regions represent the uncertainty on the  hole probabilities $\text{P}_h(d_W)$ due to the uncertainty on the effective k-values of Table \ref{T:1}.}
  \label{F:PDEPEPHCOMP}
\end{figure}
%The fact that the effective values of $\text{k}_{\text{LF}}$ and $\text{k}_{\text{VUV4}}$ are considerably smaller than $\text{k}_{\text{H2017}}$, implies that the junction average electric field of the two SiPM tested for nEXO is smaller in magnitude than the electric field of the Hamamatsu H2017 MPPC. 
An additional comparison between the three devices can be drawn analyzing their electron/hole avalanche triggering probabilities: $\text{P}_e(d_P)$ and $\text{P}_h(d_W)$, as reported in Fig. \ref{F:PDEPEPHCOMP}. The $\text{P}_e(d_P)$ of the three SiPMs saturates faster then the corresponding $\text{P}_h(d_W)$. This aspect is related to higher impact-ionization-coefficient of electrons compared to holes \cite{W.G.1972}. Additionally, Fig. \ref{F:PDEPEPHCOMP} shows that the two Hamamatsu devices have almost the same electron avalanche triggering probabilities, while those of the FBK LF are noticeably different.  $\text{P}_e(d_P)$ ($\text{P}_h(d_W)$) at fixed over voltage is always larger (smaller) for the FBK LF than for the Hamamatsu MPPCs, indicating that the FBK LF is more sensitive to UV and Vacuum Ultra Violet (VUV) wavelengths since, for these wavelengths, the avalanche mechanism is driven by electrons (Sec. \ref{S:PDEUV}).  The lower sensitivity of the Hamamatsu VUV4 MPPC in the VUV range was in fact measured in \cite{Gallina2019}. The reported saturation PDE of the Hamamatsu VUV4 and the FBK LF at an average wavelength of $189\pm7$ nm are $14.8\pm2.8$ \% and $22.8\pm4.3$ \%, respectively\footnote{The Hamamatsu VUV4 and the FBK LF were designed to cover the same spectral range while the Hamamatsu H2017 was not designed to detect VUV wavelengths.}. 
%The different spectral sensitivity of the three devices can also be analyzed by comparing their $f_e^*$ curves, as reported in Fig. \ref{F:feCOMP}. In this case, for wavelengths larger than 378 nm, the FBK LF  $f_e^*$ deviates from 1 and is always lower than the corresponding quantity for the two Hamamatsu MPPCs. This implies a hole contribution in the SiPM PDE and a partial absorption of the light past the depth of the maximum electric field.  Consequently, the FBK LF SPADs p-n junction depletion layers are closer to the SPAD surfaces and more suitable for the detection of UV and VUV wavelengths due to their short attenuation lengths. 
It is worth noting, however, that we cannot conclude that the higher efficiency of the FBK LF is exclusively due to a more optimised internal structure ($i.e.$ higher ATPs). Surface reflectivity, as well as junction depth, can also play an important role in defining the total PDE, as shown by Eq. \ref{eq:PDEmax}.

%  \begin{figure}[h]            
%  \centering
%  \includegraphics[width=0.99\linewidth]{NEW_FIGURES/MISC/fe_vs_att_All.png}
%  \caption{Fraction  of  Electron  Driven  Avalanches (EDA) ($i.e.$ $f_e^*$, derived using Eq. \ref{eq:PDE_final})  as  a  function  of  the  photon  attenuation  length for the three SiPMs analyzed in this paper. The solid lines represent fits performed with Eq. \ref{eq:fe}. The first $f_e^*$ for each SiPM is fixed at 1 (Sec. \ref{S:Summary}).}
%  \label{F:feCOMP}
%\end{figure}

\section{Conclusions}

In this paper we have presented a new analytical model to describe the SiPM PDE as a function of the reverse bias voltage. The new model was used to explain the wavelength dependence of the SiPM PDE, attributed to a combination of electron and hole avalanche triggering probabilities. In particular, we showed that the photo-generated carrier drift and diffusion in the microcell quasi neutral layers can be treated like an effective re-sizing of the microcell depletion layer boundaries, therefore increasing the effective photon collection region. The model was applied to analyze the response of three p-on-n SiPMs and can naturally be extended to any SiPM, including n-on-p devices.

\section*{Acknowledgment}
The authors would like to thank you O. Girard, G. Haefeli, A. Kuonen, L. Pescatore, O. Schneider and M. E. Stramaglia who provided the data of the characterization of the H2017 Hamamatsu detector. We would also thank: A. Pocar (University of Massachusetts, Amherst), F. Vachon (University of Sherbrooke),  M. Biroth and P. Achenbach (ICASIPM 2018), L. Doria (Institut f\"ur Kernphysik, Johannes Gutenberg-Universit\"at, Mainz), the Lumerical Device team and the nEXO collaboration for their helpful feedback on the manuscript.

\ifCLASSOPTIONcaptionsoff
  \newpage
\fi

\bibliographystyle{unsrt}
\bibliography{bibtex/bib/IEEEexample}

\end{document}